\documentclass[a4paper,11pt]{article}

\usepackage{graphicx}
\usepackage{amsfonts}
\usepackage{amsmath}
\usepackage{epstopdf}
\usepackage{psfrag}
\usepackage{fancyhdr}

\newcommand\Rey{\mbox{\textit{Re}}}  

\begin{document}

\title{
Low $R\!m$ MHD turbulence: the role of boundaries}
\author{Alban Poth\'erat\\
Applied Mathematics Research Centre, Coventry University\\
Priory Street, Coventry CV15FB, UK}

\date{1 October, 2011}

\maketitle

\begin{abstract}
In this short review, we present the main known features of MHD Turbulence 
at Low Magnetic Reynolds number, for which the flow isn't intense nor electrically conductive enough 
to disturb an externally applied magnetic field. The emphasis is deliberately 
placed on the very specific physical mechanisms of these flows, rather 
than their numerical modelling. We also focus on homogeneous magnetic fields 
which have received most attention. Since the basic properties of these 
flows have been thoroughly reviewed a number of times, this review is 
deliberately biased towards flows in bounded domains, in which the tendency to 
two-dimensionality observed in MHD flows casts the boundaries of the domain 
into a leading role.
\end{abstract}
\section{Introduction}
Magnetic fields are routinely used to attempt to control flows of electrically 
conducting fluids in metallurgical processes. The intensity of these flows 
however places them most often in a turbulent state, with direct impact on 
their dissipative, mixing and transport properties. These properties being key 
to these processes, there is a wide need to understand MHD turbulence in an 
externally applied magnetic field. In this class of flows, the magnetic 
Reynolds number is low and the flow cannot disturb an externally imposed 
magnetic field $B\mathbf e_z$. The main effect of the Lorentz force that results
 from the induced 
electric currents is then to diffuse momentum along the magnetic field lines \cite{sm82}. Flow structures tend to become elongated along this direction 
and even invariant if this effect isn't effectively opposed by inertia or 
viscosity.
In turbulent flows, 3D inertia precisely resists this phenomenon 
as it breaks up larger flows structures into smaller ones and promotes a return 
to isotropy. In an homogeneous magnetic field, turbulence is precisely 
determined by the antagonism between these two forces, and by the interaction 
parameter $N=\sigma B^2L/\rho U$ which represents their ratio ($\sigma$ and $\rho$ are the fluid's electrical conductivity and density, and $U$ and $L$ are macroscopic velocity and length scales).  Whether diffusion along magnetic field lines 
is effective enough to stretch all structures all the way between the boundaries of the domain or not decides whether the flow is 3D or in any form of 2D 
state. 

The systematic experimental investigation of MHD turbulence in liquid metals started in the 
1960's, with the works of \cite{lykoudis60,brouillette67_pf,loeffler69_pf,gardner71_jfm}. Driven by applications, these pioneering experiments   
analysed turbulence in ducts and pipes, rather than "homogeneous" turbulence 
and were reviewed in detail by Lykoudis \cite{lielausis75}.
In this review, we focus on the physical properties of turbulence with no mean 
flow, as first investigated experimentally by \cite{alemany79}. 
We shall underline the influence of the walls not only on the 2D states of MHD 
turbulence but also on the crutial intermediate states that exist between 
two-dimensionality and full three-dimensionality.
 We shall first explore the consequences of the antagonism between 
inertia and the Lorentz force  on turbulence far 
from walls (section \ref{sec:far}). Section \ref{sec:2dtrans} will be dedicated to the mechanisms of the transition between 3D and strictly 2D turbulence. 
In section \ref{sec:q2d}, we shall review the possible states of quasi-2D turbulence in the presence of Hartmann walls, while section \ref{sec:appearance} will underline some of the author's recent results on the appearance of 
three-dimensionality in quasi-2D MHD flows.
\section{MHD turbulence far from walls}
\label{sec:far}
The most generic mechanisms of MHD turbulence are best singled out in flow 
regions far from boundaries where the Lorentz force is at least strong enough 
to balance inertia, \textit{i.e.} $N\simeq1$ or $N>>1$.
Despite the presence of a strong magnetic field, it is usually 
acknowledged that 
the three ranges of 3D homogeneous non-MHD turbulence still exist: the flow is 
usually assumed forced at some large scales, whose non-universal behaviour is 
dictated by the particular forcing and the boundary conditions. Energy is 
cascaded down along the inertial range, which by contrast with the large 
scales is believed to exhibit a somewhat universal behaviour. The inertial 
range stops at 
the "small scales" where viscous friction becomes dominant. Unlike in non-MHD
 turbulence at high Reynolds number, however, the Joule dissipation incurred 
by the Lorentz force
 extracts energy at all scales so that not all the energy pumped out of 
the large scales survives along the inertial range, which therefore exhibits a 
steeper energy spectrum than the $E(k)\sim k^{-5/3}$ law of homogeneous 
hydrodynamic turbulence ($k$ stands for 
the usual wavenumber used to identify vortex size) \cite{k41}. 
The dynamics of the inertial range of MHD turbulence are usually described 
using two assumptions \cite{alemany79}:
1) At each scale, inertial forces balance Lorentz forces and 
%
2) Anisotropy remains the same at all scales, over the inertial range. 
%
These lead to the scalings for the power spectral density, and 
for a "geometric anisotropy":
%
\begin{eqnarray}
E(k_\perp) \sim U_0^2k_{\perp}^{-3}
\label{eq:k-3}\\
\frac{k_z}{k_\perp} \sim \frac{Re^{1/2}}{Ha}=N^{-1/2}
\label{eq:anis_h}
\end{eqnarray}
$Re$ is a Reynolds number based on the large scale velocity $U$ and length $L$,
 the ratio $\frac{Ha^2}{Re}=N$ is the corresponding interaction parameter, and 
the square of the Hartmann number $Ha=LB(\sigma/(\rho\nu))^{1/2}$ represents 
the large scale ratio of the Lorentz to viscous forces.
The subscript $\perp$ stands for components of vectors orthogonal to 
$\mathbf B$, assumed aligned with $\mathbf e_z$.
 Unlike the Lorentz force, viscous friction is only effective at very small scale. When active, it stops the energy cascade so that the smallest scales  
are heuristically defined as the smallest possible structures of the inertial 
range which are not destroyed by viscosity. This leads to
%
%
%
\begin{eqnarray}
k_{\perp_{max}} \sim Re^\frac{1}{2}
\label{eq:kp_max},  \qquad&
k_{z_{max}} \sim \frac{Re}{Ha}.
\label{eq:kz_max}
\end{eqnarray}
The corresponding number of degrees of freedom of the flow 
scales as 
$N_f \sim k_{\perp_{max}}^2k_{z_{max}} \sim \frac{Re^2}{Ha}$. These scalings 
reflect that Joule dissipation strongly reduces the number of degree of 
freedom, since $N_f\sim Re^{9/4}$ in non-MHD turbulence. In other words, 
the lesser amount of energy that survives the journey down the inertial 
range is dissipated by viscous friction at much larger "small scales" than in 
non-MHD turbulence.\\

\begin{figure}
\centering
\includegraphics[width=7cm]{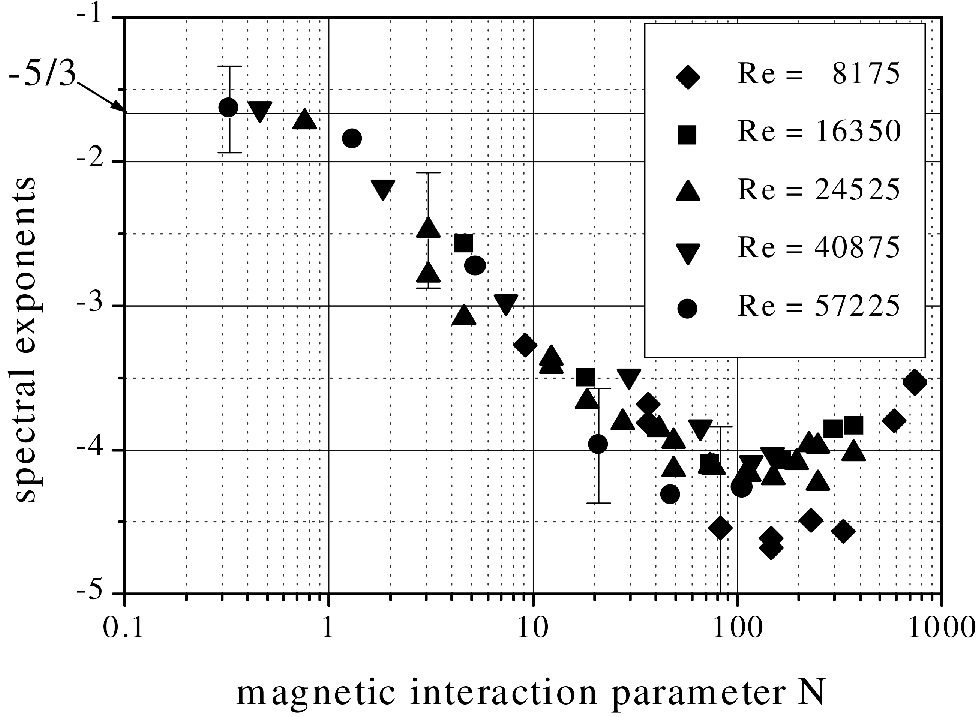}
\caption{Variations with $N$ of the exponent in the scaling law for the 
inertial range of the energy spectrum found by \cite{eckert01_ijhff} in the 
turbulent flow in the wake of a fixed grid in a rectangular channel.}
\label{fig:exponents}
\end{figure}

The validity of these simple scalings has been tested experimentally in several occasions, in particular the $k^{-3}$ law. \cite{kolesnikov74} and \cite{alemany79} measured turbulent spectra in a turbulent flow of liquid metal, either 
subjected to an homogeneous magnetic field or to no field. They were  
able to convincingly recover both the $E(k)\sim k^{-5/3}$ law in non-MHD case, 
and the $E(k)\sim k^{-3}$ law with an applied magnetic field corresponding to 
an interaction parameter of at least a few units. In both cases, the 
experiments were designed in such a way as to eliminate 
any significant influence from the walls.  More recently, \cite{eckert01_ijhff}
measured turbulent spectra in a rectangular channel, to find that the spectral exponent in the inertial range depended  non-monotonously on the 
value of $N$, with values in the range $[-5,-5/3]$ (see figure \ref{fig:exponents}). Two main explanations 
were put forward: firstly, it was deemed possible that for $N>1$, 
the larger structures, at least, could extend across the whole channel. Their 
behaviour would then be governed by 2D turbulence leaving the flow in a partly 2D, partly 3D state.
For exponents lower than -3, \cite{branover94,branover99}'s theory was invoked,
 as it suggests that the presence of helicity in the flow can lead to such 
steep spectra. Although no definite evidence for this explanation could be 
put forward, it does find support in the fact that the presence of large 2D 
structures in the vicinity of walls can indeed generate helicity by Ekman 
pumping (see section \ref{sec:q2d}).\\
On the more theoretical side, attempts were made to justify these heuristic 
scalings either numerically of by analysing the dynamical system associated 
to a generic turbulent flow.
Rigorous estimates of the dimension of its attractor were derived by 
\cite{pa03,pa06_pf} that indeed confirmed the exponent of $Ha$ in 
these scalings, without the need for any additional assumption than those the 
Navier Stokes equations rely on.\\
Since the 80's, the properties of Low-Rm MHD turbulence have been to a large 
part analysed numerically, in periodic domains, with various large scale 
forcings. These aspects are reviewed in detail in \cite{knaepen08_arfm}. 
Among these studies, \cite{vorobev05} raised the question of the validity of 
the assumption that anisotropy is constant across the inertial range. They 
stressed that anisotropy could be defined in a number of ways, 
and were able to show that for $N>1$, both the anisotropy of the 
velocity gradients and the kinematic anisotropy (defined as the ratio of 
energies along and across the field direction) were reasonably 
scale-independent in the inertial range, for several types of forcing.\\
The picture can be refined by inspecting how the energy is distributed in the 
$(k_\perp,k_z)$ spectral space.  In this regard, it was early recognised that 
the selective nature of Joule dissipation severely damped modes within the 
"Joule cone", defined as the region of spectral space such that  
$k_z/k_\perp < N^{-1/2}$ \cite{moffatt67}. Consequently, hardly any 
energy remains there unless it is maintained by external forcing. 
Experiments by \cite{caperan85} 
showed that  because of their finite number, energy containing modes were in 
fact localised in a torus in $(\mathbf k_\perp,k_z)$, whose pointy inward edge coincided with the Joule Cone. This was later confirmed by the numerical 
simulations of 
\cite{burattini08_pf,pdy10_jfm},  who showed that the radial section of the 
torus was shaped as a cardioid curve, and that spectral energy transfer essentially 
occurred through surfaces of the same family. \cite{pdy10_jfm} 
found that the lines of constant energy tended to coincide with those of the 
decay rate due to combined viscous and Joule friction 
$\lambda_{\mathbf k}=k^2+Ha^2 k_z/k^2$. 
They noted that the anisotropy of low-$R\!m$ MHD turbulence was naturally 
rendered by the sequence of scalar decay rates $(\lambda_{\mathbf k})$. This 
lifted the need for separate laws along and across the field (\ref{eq:kp_max}),
 which could be replaced by a single one involving the forcing scale $k_f$:
\begin{equation}
\frac{\sqrt{|\lambda^{\rm max}|}}{2\pi k_f} \simeq 0.5 \Rey^{1/2}.
\label{eq:lambda_max}
\end{equation}
\begin{figure}
\centering
\includegraphics[width=7cm]{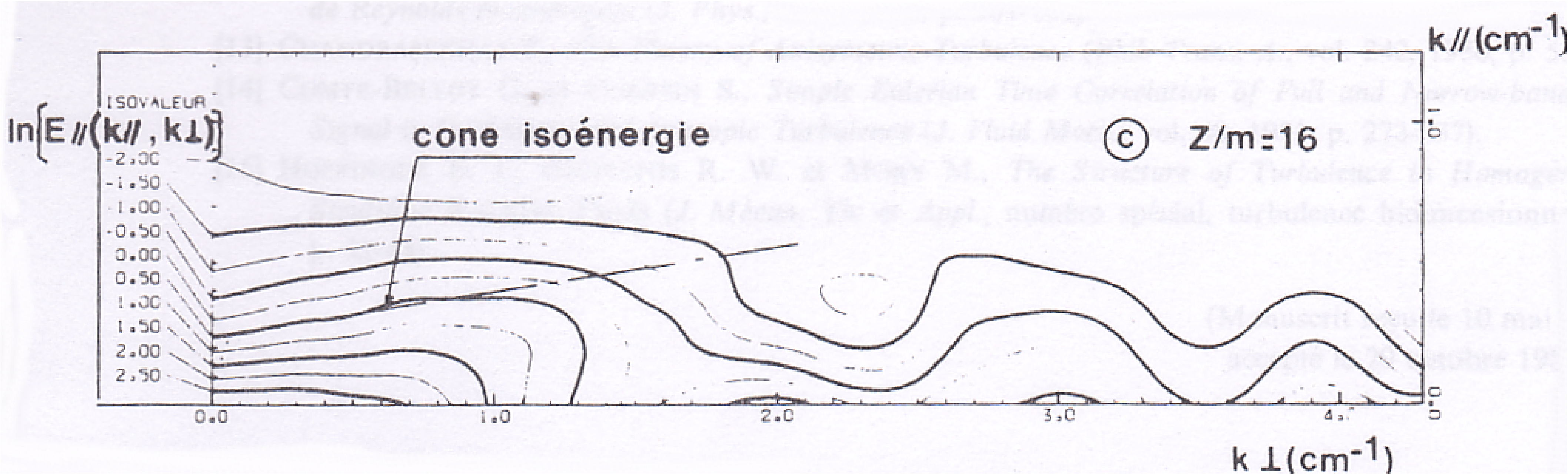}
\psfrag{x}{$k_x$}
\psfrag{y}{$k_y$}
\includegraphics[width=5cm]{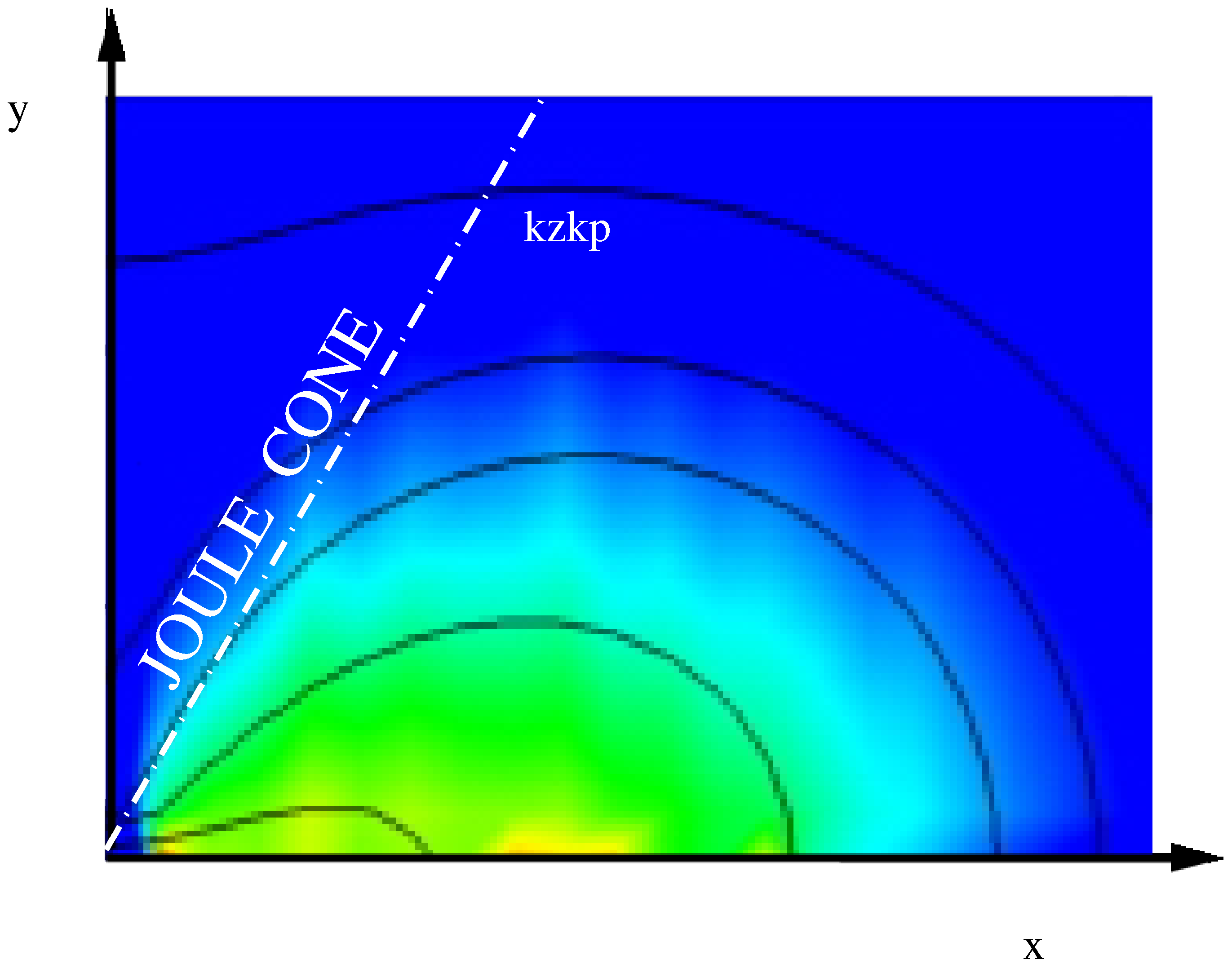}
\caption{Contours of turbulent kinetic energy, as observed in the wake of a 
moving grid by \cite{caperan85} and in direct numerical simulations of 
statistically forced 3D turbulence \cite{pdy10_jfm}. In both figures, 
most of the energy is contained outside the Joule cone, within a cardioid-shaped region of the spectral space.}
\label{fig:2dspectra}
\end{figure}
\section{Transition to and from strict two-dimensionality}
\label{sec:2dtrans}
One of the most distinctive features of MHD turbulence is its tendency to 
become 2D, highlighted in introduction. But the question of just 
how close to two-dimensionality MHD turbulence can become cannot find an 
answer without specifying the boundary conditions of the problem, at least 
along those 
boundaries that intercept the magnetic field lines. In his seminal 1967 paper 
\cite{moffatt67}, Moffatt showed that in a freely decaying unbounded flow, the 
anisotropy of structures increased indefinitely, and that the flow tended 
towards a limit  state where the ratio of kinetic energies along to across 
the magnetic field direction would be of 1/2. 
This spread a vision of 2D MHD turbulence as a limit state rather than as an 
achievable flow. 
Later, numerical simulations of freely decaying MHD turbulence in a spatially
periodic domain by \cite{schumann76} indeed exhibited strictly 2D structures when 
the initial interaction parameter was above $50$. Here, strict 
two-dimensionality was only made possible by the finite length imposed by 
periodic boundary conditions chosen along the magnetic field lines. Similar 
observations were more recently made in forced MHD flows \cite{zikanov98}, 
still in 3D periodic domains.\\
The reverse transition that leads three-dimensionality to appear in an initially
 strictly 2D flow has received less attention. 
\cite{zikanov98} mention the existence of intermittent regimes 
in forced flows. More recently, \cite{boeck07} and \cite{thess07_jfm} argued 
that such intermittency could appear in regimes where diffusion of momentum 
along the field direction by the Lorentz force was sufficiently strong to 
create 2D structures but not sufficiently dominant to prevent the 
development of 3D instabilities that disrupted these 
structures. Indeed, without dissipation at the boundaries, the flow can become
 strictly 2D, a state in which the Lorentz force vanishes 
completely, leaving the growth of 3D perturbations unimpeded. 
These perturbations break down  2D structures to restore a 3D state in 
which the Lorentz force starts acting again. 
This dynamical instability can produce an intermittent behaviour in domains 
with periodic or no-slip boundaries, but this effect has never been observed in 
domains bounded by no-slip walls, where strong dissipation is always present in wall boundary layers.\\
When strict two-dimensionality is achieved, the electric current density and 
the Lorentz force vanish entirely so the flow strictly recovers the properties 
of non-MHD 2D turbulence: above the injection scale, energy is cascaded 
upwards up to large coherent structures whose dynamics is dictated by the 
forcing and the conditions along boundaries parallel to $\mathbf B$. The 
corresponding power density spectrum exhibit a $k^{-5/3}$ slope.  Below this 
scale, enstrophy is cascaded along a $k^{-3}$ energy spectrum, down to the 
Kraichnan scales $k_{\perp_{max}}\sim Re^{1/2}$ (reviews of 2D turbulence can 
be found in \cite{tabeling02_pr} and \cite{clerx09_amr}).
\section{Quasi-2D MHD turbulence}
\label{sec:q2d}
\begin{figure}
\centering
\includegraphics[width=5.75cm]{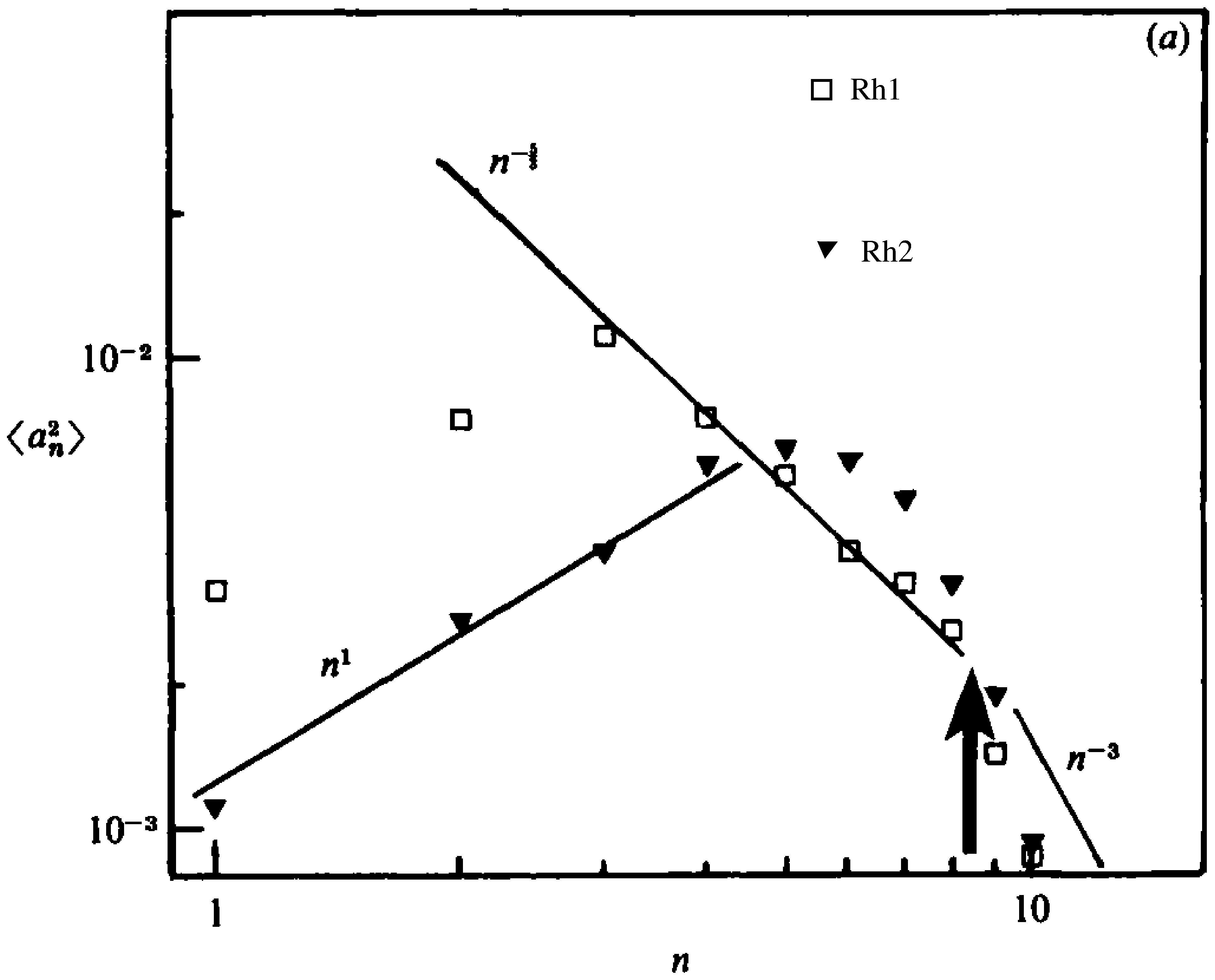}
\includegraphics[width=6.5cm]{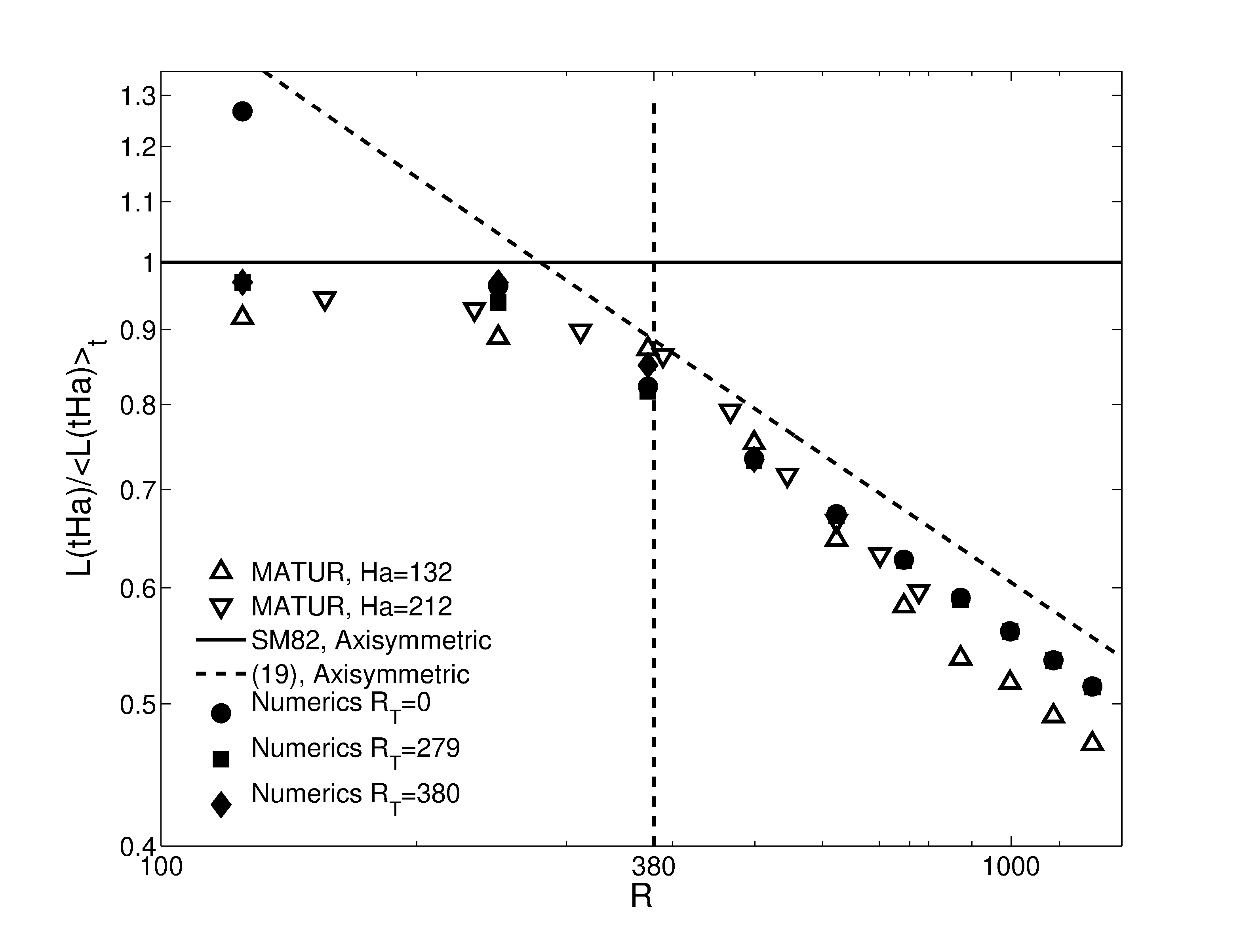}
\caption{Left: Kinetic energy spectra measured by \cite{sommeria86} in a 
quasi-2D turbulent layer of mercury pervaded by a transverse magnetic field. The inverse energy cascade is identified through the $k^{-5/3}$ slope. 
Right: variation of total angular momentum against 
$R=Re/Ha$, for the quasi-2D annular flow 
electrically driven in the MATUR experiment (experiments from 
\cite{messadek02_jfm} referred to as "MATUR", numerical simulations from 
\cite{ps11_pf}, based on \cite{alboussiere00_pf}'s model for the turbulent Hartmann layer). Values are normalised by the value of the angular momentum 
predicted on the assumption of a laminar Hartmann layer: the change in slope 
reflects the intensification of friction when the layers become turbulent.}
\label{fig:q2dturbulence}
\end{figure}

Strictly and intermittently 2D flows have only ever been achieved 
in numerical simulations with either periodic or free-slip boundary conditions, but never in a laboratory where the influence of walls and wall friction can 
never be completely avoided. As such, no-slip walls are an intrinsic part of 
MHD turbulence and become especially important in flow regimes that approach 
the 2D state.\\ 
The most obvious feature of wall-bounded flows is the presence of 
\emph{Hartmann boundary layers} along boundaries that intercept the magnetic 
field lines. In these layers, of thickness $\sim Ha^{-1}$, viscous friction opposes the Lorentz force to 
maintain a velocity gradient along $\mathbf B$ \cite{moreau90}. Strict 
two-dimensionality is thus only possible outside of these layers and the 
corresponding flows are only \emph{quasi-2D}, rather than strictly 2D. The 
influence of the walls in quasi-2D MHD was first analysed by \cite{sm82}, who 
showed that in the limit of large $N$ and $Ha$, inertia was negligible in the 
Hartmann layers and the flow in the core was not only dynamically 2D 
(\emph{i.e.} $\mathbf u\cdot \mathbf B=0$), but also kinematically 2D 
(\emph{i.e.} $\partial_\mathbf B  \mathbf u=0$). The absence of inertia in the 
layers doesn't prevent the development of 2D turbulence in the core, but the 
Hartmann layers exert linear friction on it so that a flow confined within a 
channel orthogonal to the field is described by a shallow water equation of 
the (dimensional) form:
\begin{equation}
\frac{d\mathbf u_\perp}{dt}+\nabla_\perp p= \nu \nabla_\perp^2 \mathbf u_\perp -\frac{\mathbf u_\perp}{t_H}+\mathbf f, 
\end{equation}
where $\mathbf f$ represent an externally applied force density.
For a given fluid, the linear damping time $t_H=H^2/\nu Ha^{-1}$ is controlled 
by the intensity of the magnetic field and the channel depth $H$. In this same 
channel configuration, \cite{kolesnikov74}  and \cite{sommeria86} 
experimentally showed that the dynamics of such flows was essentially that of 
2D turbulence and the latter work presents an evidence of an inverse energy 
cascade (see figure \ref{fig:q2dturbulence}). Unlike in 2D turbulence 
however, linear friction introduces an energy sink at scale 
$k_\perp^{min}\sim Ha/Re$, as structures larger than the corresponding size 
cannot survive the action of friction. If this 
largest possible scale is smaller than the typical dimension of the domain, 
Hartmann friction prevents the condensation of energy in modes dictated by the 
boundary conditions and stops the possible energy pile-up in the large scales
associated with this phenomenon \cite{tabeling02_pr}.\\
In the regimes of moderately high inertia where $N\sim1$, rotation at the 
scale of individual vortices becomes strong enough to drive local poloidal 
recirculations, through a local \emph{Ekman pumping} 
mechanism \cite{psm00}, an effect that is well known in rotating flows 
\cite{pedlosky87}.
 These secondary flows were shown to alter the properties of quasi-2D 
turbulence, by introducing non-linear anisotropic diffusion along the flow 
streamlines, that tends to damp small scale fluctuations \cite{dellar04,psm05}. 
Local Ekman pumping is also a source of helicity, which, according to 
\cite{branover94} induces steep turbulent spectra. This too, is an expression 
of the damping of small scales.\\
The more recent experiments of \cite{messadek02_jfm} exhibited a further 
interesting regime of MHD turbulence in a channel where the core remained 
2D, as in \cite{sommeria86}'s experiments, but where the Hartmann boundary 
layers were turbulent. This happened whenever $Re/Ha> 380$ and $N>>1$. Unlike 
the friction exerted by laminar Hartmann layers, that due to 
turbulent Hartmann layers varies non-linearly with the core velocity 
(see figure \ref{fig:q2dturbulence}). It incurs a much higher global 
dissipation, alters the scaling of the large scales \cite{ps11_pf}, and 
most likely the rest of the spectrum, all the way down to the size of the 
small scales.
\section{Appearance of three-dimensionality}
\label{sec:appearance}
Although dominant in the quasi-2D state of MHD turbulence, the role played by 
boundaries isn't confined to this regime. \cite{sm82} argued that since the
anisotropy of a given structure of size $(l_\perp, lz)$ resulted from a 
competition between diffusion of momentum along the magnetic field
and return to isotropy driven by inertia, then in a channel of width $H$, a 
critical size $l_\perp^{2D}$ existed above which structures were quasi-2D and 
below which they were 3D:
\begin{equation}
\frac{l_\perp^{2D}}{H}<\left(\frac{\rho U}{\sigma B^2 H}\right)^{1/2}.
\label{eq:l2d}
\end{equation}
This remarkable property was verified experimentally only recently when 
\cite{kp10_prl} forced MHD turbulence in a cubic, insulating container placed 
in an homogeneous magnetic field. By comparing the frequency spectra derived 
from the electric potential gradients measured near both Hartmann walls, at 
opposite locations, they found a cutoff frequency $f_c$ separating 2D from 3D 
fluctuations:
\begin{equation}
f_c\simeq 1.7 \tau_{u^\prime}^{-1} {\rm N}_t ^{0.67}.
\label{eq:l2d}
\end{equation}
The true interaction parameter $N_t=N(H/L_f)^2$ was based on the scale at which 
the flow was forced $L_f$ and the turnover time $\tau_{u^\prime}$ was 
associated to RMS of velocity fluctuations.
This experiment also singled out further mechanisms at play when the Lorentz 
force wasn't strong enough to achieve quasi-two dimensionality in forced, 
established flows: at high $Ha$, 
the flow was quasi-2D as in \cite{sommeria86}'s experiment. For slightly lower 
values of $Ha$, a form of three-dimensionality, called \emph{weak} was 
observed where flows in planes orthogonal to the field were topologically 
identical but of intensity decreasing with the distance to the Hartmann wall 
where the forcing was applied. This observation recovers the theoretical and 
numerical predictions of \cite{psm00, muck00}. These authors proved indeed
 that 2D inertia induced electric eddy currents between Hartmann layers
and the bulk, that caused differential rotation and led columnar vortices to
assume a 3D \emph{barrel}-like shape. Weak three-dimensionality is therefore
a direct consequence of the presence of Hartmann walls.\\
\begin{figure}
\centering
\includegraphics[width=8cm]{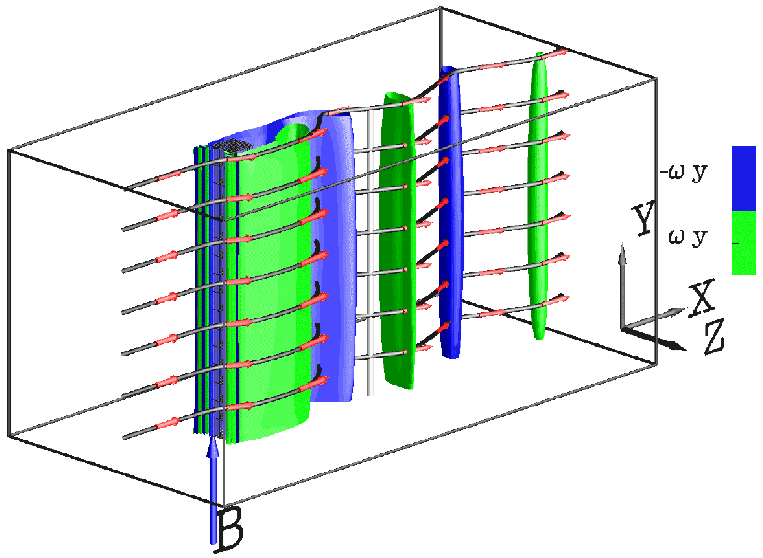}
\caption{Barrel-shaped vortices found in the wake of a rectangular cylinder, according to the numerical simulations of \cite{muck00}.}
\label{fig:barrel}
\end{figure}
At moderate values of $Ha$, partial vortex merging was  
observed where vortices generated near one Hartmann wall were elongated along 
the magnetic field and merged near the opposite Hartmann wall, leading 
to Y-shaped vortices. At moderately high $Ha$, vortex pairing 
was unsteady, but most remarkably, for the lowest values of $Ha$, this 
phenomenon could lead to a re-stabilisation of the flow with steady Y-shaped 
vortices. Although the flow clearly wasn't turbulent in these regimes, these 
findings single out some of the mechanisms of the antagonism between momentum 
diffusion along the field lines and inertia, that give birth to remarkable 
flow structures directly relevant to dynamics of MHD turbulence around the 
2D-3D transition.\\
\begin{figure}[h!]
\centering
\includegraphics[width=4cm]{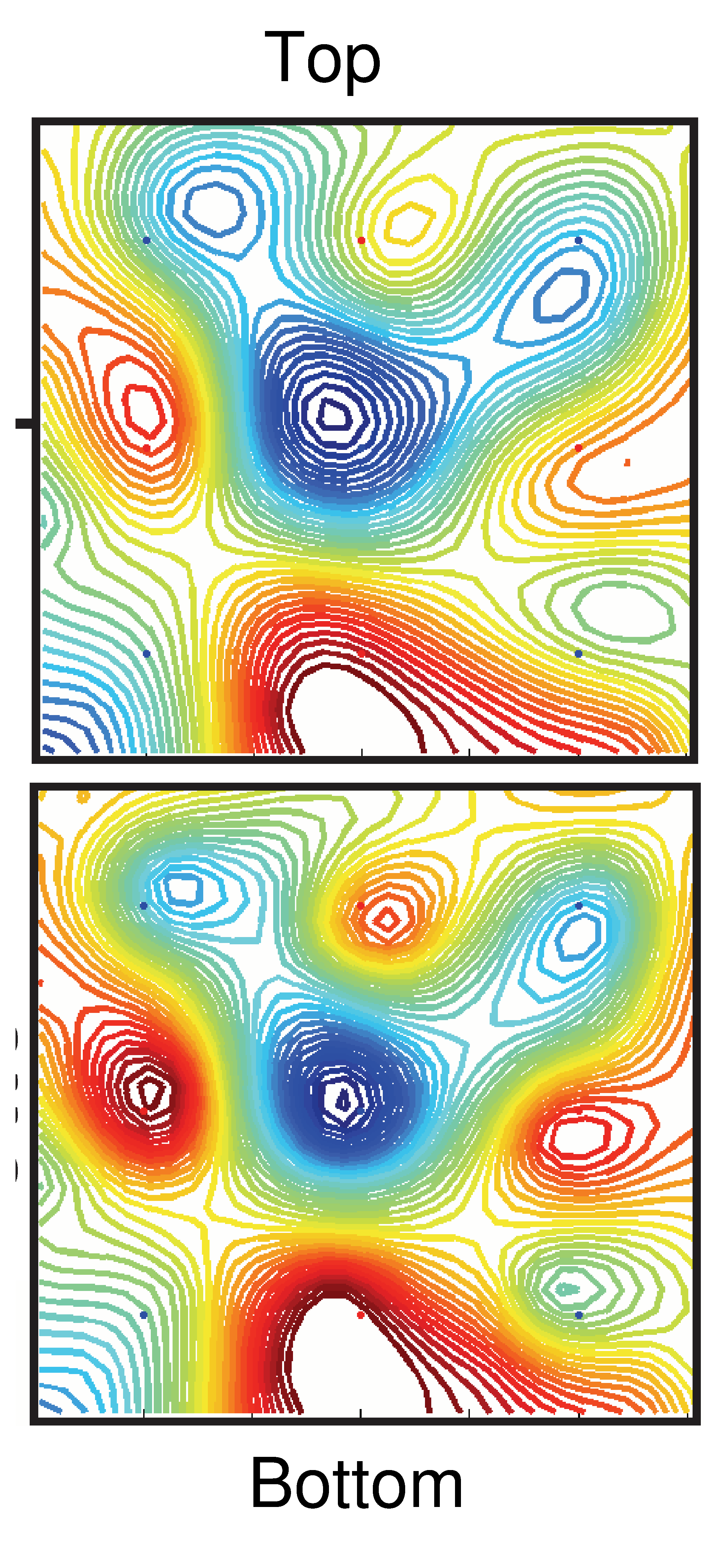}
\includegraphics[width=4cm]{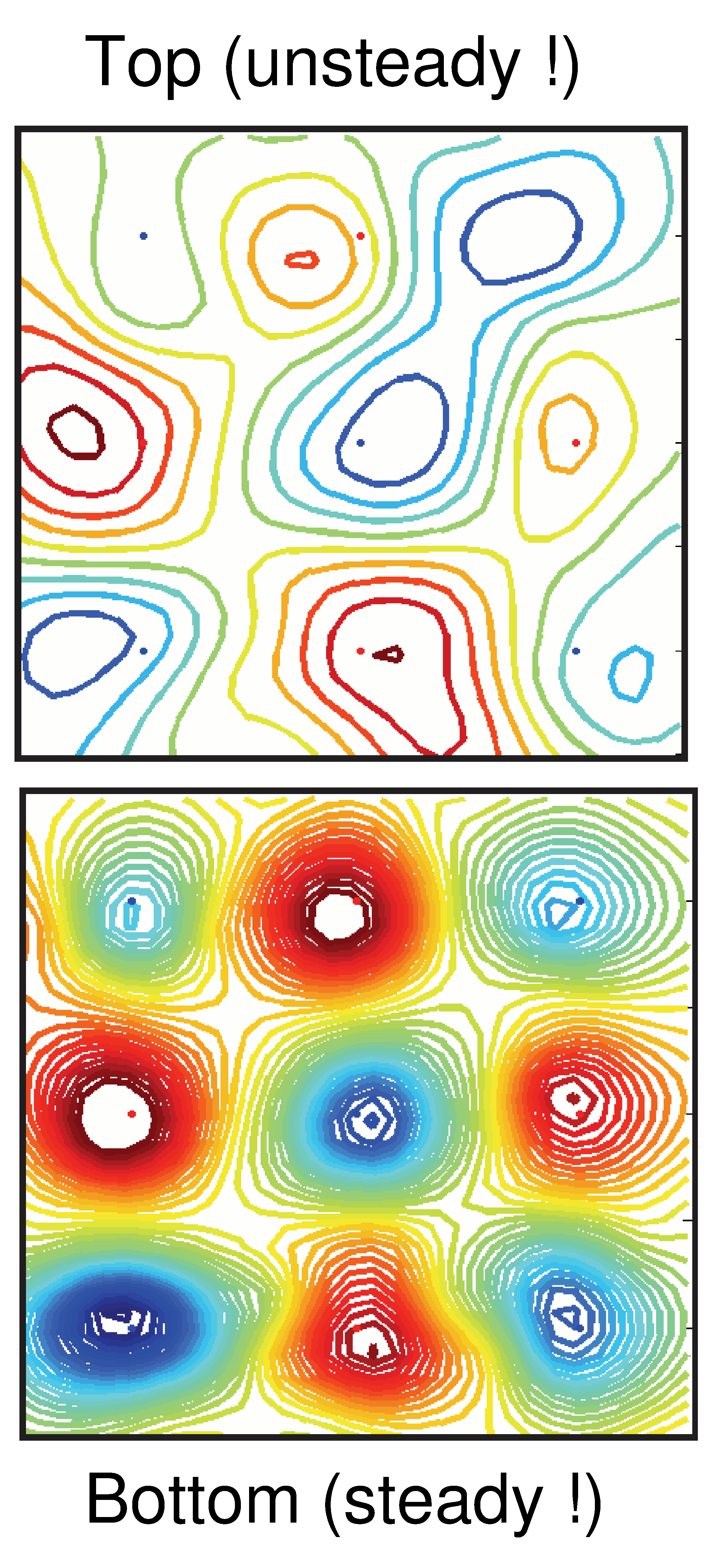}
\includegraphics[width=4cm]{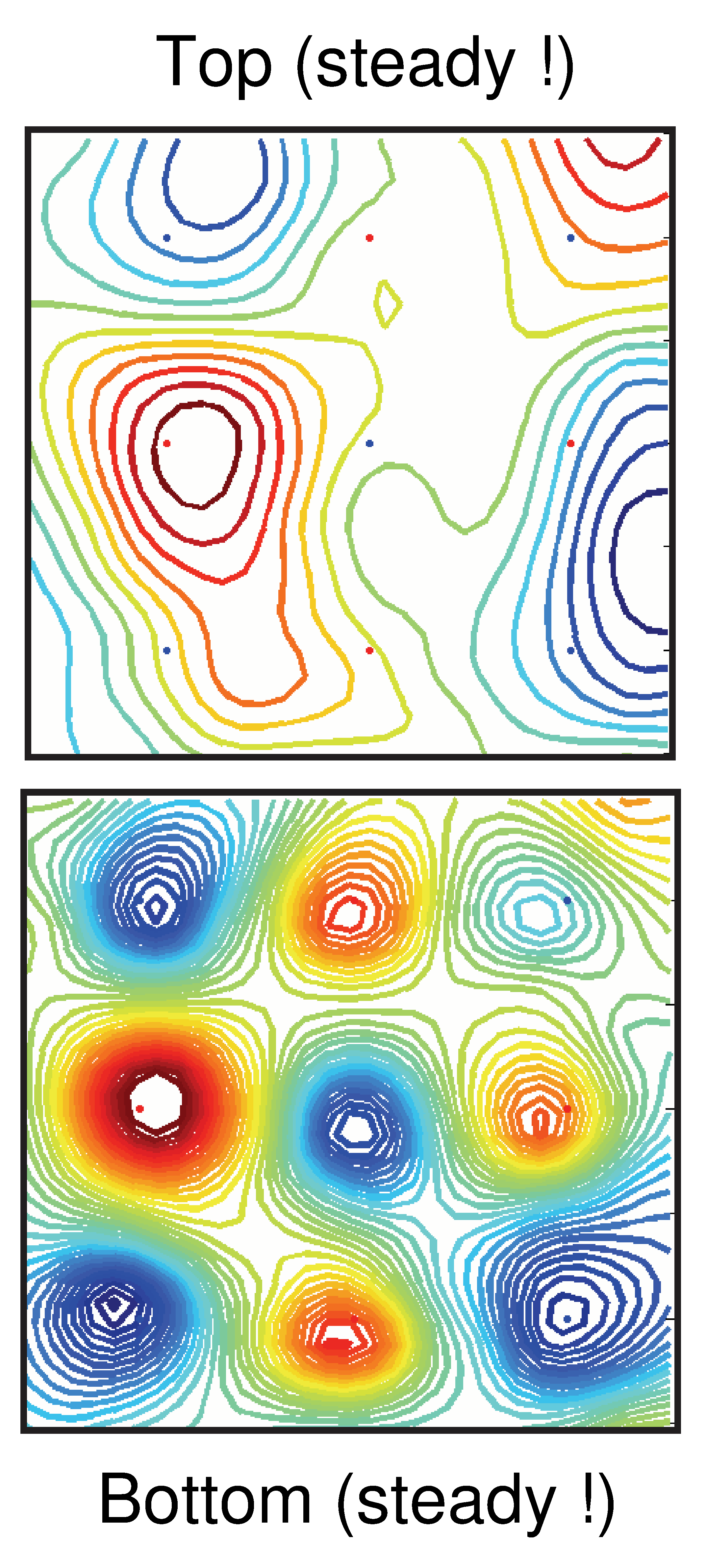}
\caption{Snapshots-contours of electric potential measured on two Hartmann 
walls facing each other \cite{kp10_prl}. The flow is forced by injecting 
electric current through a square array of electrodes embedded in one of the 
walls (denoted "bottom"). Left: $Ha=18220$, 
$Re\simeq10^4$, contours on both Hartmann walls are practically identical so 
the flow is quasi-2D turbulent. Middle: $Ha=1822$, $Re=409$, 
the flow is steady near the bottom wall but periodic merging appears near the 
top wall. Right: $Ha=1822$, $Re=512$, the flow is steady, but vortices are 
not columnar anymore and exhibit 3D reconnexions.}
\label{fig:kp10}
\end{figure}
\section{Conclusion}
To conclude this short review, understanding Low-$Rm$ MHD turbulence 
is still very much a task in progress, particularly in wall-bounded 
configurations or more realistic ones. If some of the basic mechanisms are now 
well understood, at least heuristically, hardly any exact result is  
available for this rather specific type of turbulence (such as as 4/5th law in 
homogeneous turbulence). The most distinctive feature of MHD turbulence is 
possibly its tendency to two-dimensionality, which, unlike in turbulent 
flows in rotation, incurs strong dissipation. In this regard, the 
conditions of the transition between quasi-2D and 3D turbulence are still very 
poorly understood. Recent progress indicate that the nature of the boundaries 
play a lead role in it: while strict two-dimensionality is only possible in 
domains bounded by non-dissipative boundaries, the presence of no-slip walls
implies that three-dimensionality appears progressively in the 
flow, rather than because of the instability of quasi-2D structures.
Three-dimensional instabilities still probably occur, but develop in states 
where turbulence may already exhibit several possible forms of 
three-dimensionality.  Transition to three-dimensionality along such a route 
is to this day unexplored, and is likely to differ significantly from that 
found in simulations without dissipative boundaries.
%

\end{document}